\documentclass{aa}     
\usepackage{epsfig}
\begin{document}
\thesaurus{03 (11.01.2; 11.09.1: NGC 2992; 11.11.1; 11.19.1; 11.19.2) }

\title{A kinematical analysis of NGC 2992. \thanks{Based on data
    obtained at the European Southern Observatory, La Silla, Chile}}

\author{
  I. M\'arquez \inst{1,2}
\and
  C. Boisson \inst{3}
\and
  F. Durret \inst{2,3}
\and
  P. Petitjean \inst{2,3}  
}
\offprints{I. M\'arquez, IAA, Spain}
\institute{
        Instituto de Astrof\'\i sica de Andaluc\'\i a (C.S.I.C.), 
Apartado 3004 , E-18080 Granada, Spain
\and
        Institut d'Astrophysique de Paris, CNRS, Universit\'e Pierre et 
Marie Curie, 98bis Bd Arago, F-75014 Paris, France 
\and 
        DAEC, Observatoire de Paris, Universit\'e Paris VII, CNRS (UA 173), 
F-92195 Meudon Cedex, France 
}
\date{Received July 14, 1997 ; accepted, 1998}

\maketitle
\markboth{M\'arquez et al.: A kinematical analysis of NGC 2992}{}

\begin{abstract}
We present long slit spectroscopy for the [OIII]  and H$\alpha$ wavelength 
ranges along nine different position angles for the Sa Seyfert 1.9 galaxy 
NGC 2992. Double profiles are present in several regions, suggesting that the 
gas is not simply following galaxy rotation. A simple kinematical model, which 
takes into account circular rotation together with a constant radial outflow, 
seems to be a good approximation to account for the observed kinematics. 
\keywords{galaxies: active - galaxies: individual: NGC 2992 - 
galaxies: kinematics and dynamics - galaxies: Seyfert - galaxies: spiral}
\end{abstract}

\section{Introduction}

Disturbed morphologies of ionized nebulosities surrounding active
galactic nuclei are frequently observed, mainly in high ionization
gas.  The morphology is usually interpreted in terms of a conical
or biconical shape centered on the nucleus. The spectacular ionizing
cone discovered in NGC 5252 by Tadhunter \& Tsvetanov (1989) is
thought to arise from interstellar matter lit up by radiation from the
nuclear non stellar continuum escaping the central regions
through the hole of an obscuring torus surrounding the nucleus. The
situation is far less clear in the other reported cases where the cone
may be seen only on one side of the nucleus or has less marked edges
(cf Wilson \& Tsvetanov 1994 for a review). The possibility of having
outflows or inflows in cones centered on the nucleus has been invoked
to account for the kinematical properties of several Seyferts
(e.g. Wilson et al. 1985), but kinematics have been studied in detail
only in a few of the Seyferts which show evidence for galactic
outflows.

In this paper we present new results on the kinematics of the ionized
gas in the highly inclined ($i$=70$^\circ$) Sa Seyfert galaxy
NGC 2992, crossed by a disturbed dust lane oriented along the major
axis ($\phi$=15$^\circ$, RC3 catalogue). This edge-on galaxy is
connected by a tidal tail to a close companion, NGC 2993, at a
projected distance of 2.9 arcmin (35.5 kpc for
H$_0$=50 km s$^{-1}$ Mpc$^{-1}$) and with a 109 km s$^{-1}$ velocity
difference (see RC3 catalog), which may well have important perturbing
effects on its dynamics. Combined optical broad and narrow band images
reveal a complex structure (Durret \& Bergeron 1987, Wehrle \& Morris
1988). The [OIII] and H$\alpha$ images show an arc of emission
southeast of the nucleus (which could be interpreted as HII regions in
the spiral arm) as well as a finger of emission emerging from the
northwest portion of the nucleus, pointing northward.  At 20cm radio
wavelength, NGC 2992 reveals a radio source of total extent 25 arcsec and
major axis PA$\sim 160^\circ$, with a one--sided extension along
PA$\sim 130^\circ$ (Ward et al, 1980; Hummel et al., 1983). At smaller
scale an 8-shaped structure, with the nucleus at the crossing-point,
is visible at 6cm along PA=160$^\circ$ and interpreted as
limb-brightened bubbles or magnetic arches (Ulvestad \& Wilson 1984;
Wehrle \& Morris 1988). A comparison of the line emission images with
the radio images shows no correlation.

Colina et al. (1987) have mapped NGC 2992 by using long slit
spectroscopy roughly along the major axis and the two axes given by
the high resolution radio map by Ulvestad \& Wilson (1984). From their
kinematical data in the [OIII] lines, they observe blue asymmetric
profiles only in the very center of the galaxy. They find that the
nuclear and off-nuclear regions are dynamically decoupled, and suggest
that there are non-circular motions due to radial flows and tidal
interaction with NGC 2993. They interpret their data as being
consistent with a radial outflow of gas in a plane which is not
coaligned with the galactic plane, rather than outflow within a cone.
On the contrary, Tsvetanov et al.  (1995) describe NGC 2992 as a good
candidate for having large-scale minor axis outflows with velocities
up to 200 km s$^{-1}$.

Since NGC 2992 is a highly inclined object, it is a good candidate to
sample gas motions out of the disk plane. In the following, we present
a first order kinematic model of the ionized gas in NGC 2992, based on
a set of long slit spectra.

\section{The data }\label{data}

\begin{table}
\caption[ ]{Journal of Observations. }
\begin{footnotesize}
\begin{tabular} {c l r c c c}
\\
\hline
\\
Observing & PA/offset & Exp. &  Wavelength & Seeing\\
date      &           & time (s) &  range (nm) &  (")\\
\\
\hline
\\
12/3/88 &  17/0    & 2700 &  625--725 & $\leq$ 1.2\\
"       & 100/0    & 3600 &  "        & "\\
"       & 150/0    & 2700 &  "        & "\\
"       &  58/0    & 3600 &  "        & "\\
13/3/91 & 122/0    & 7200 &  405--600 &  1.5\\
"       &  30/0    & 3600 &  "        & "\\
14/3/91 & 122/0    &  600 &  "        & 1.1\\
        & 122/4.6NE& 7200 &  "        & "\\
        &  30/0    & 3600 &  "        & "\\
15/3/91 & 122/0    &  600 &  "        & 1.8\\
        & 122/6.8NE& 7200 &  "        & "\\
        &  80/0    & 7200 &  "        & "\\
16/3/91 & 122/0    &  600 &  568--746 & 1.5\\
        & 122/4.6NE& 3600 &  "        & "\\
17/3/91 &  58/0    & 3600 &  407--601 & 1.3\\
        & 150/0    & 3600 &  "        & "\\
        & 100/0    & 3600 &  "        & "\\
\\
\hline
\end{tabular}
\end{footnotesize}
\protect\label{datatab}
\end{table}

Long-slit spectroscopy was obtained during two runs with the ESO 2.2m
telescope in 1988 and with the ESO 1.52m telescope in 1991 using a
Boller \& Chivens spectrograph. Details on the observations are given
in Table \ref{datatab}. For the 1988 run, the CCD used was \#8, with a
pixel size of 15 $\mu$m, corresponding to 0.9 arcsec on the sky; the
grating was \#26 (dispersion 59.5 \AA /mm), giving a spectral
resolution of 1.9 \AA .  For the 1991 run, the CCD used was \#13, with
a pixel size of 15 $\mu$m, corresponding to 0.68 arcsec on the sky; grating
\#23 was used (dispersion of 129 \AA /mm), giving a spectral
resolution of 4.8 \AA . The slit width was 2 arcsec for both runs.
Wavelength calibration lamps (He-Ar) were taken just before or just
after each exposure.

\begin{figure}
\vbox to 0.5cm{}
\centerline{\psfig{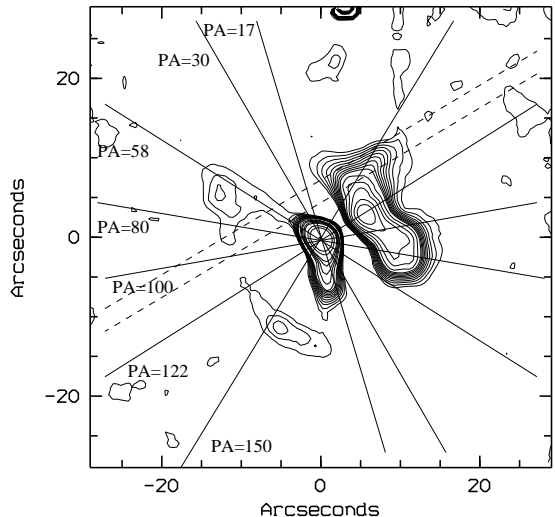}}
\caption[ ]{Isophotes of the [OIII] narrow band image from Durret \&
Bergeron (1987) with the slit positions superimposed. North is to the top
and east to the left.}
\protect\label{slits}
\end{figure}

A map of the [OIII] narrow band image from Durret \& Bergeron (1987),
with the slit positions superimposed, is displayed in
Fig. \ref{slits}.  The exact positions of the slits on this image are
derived assuming that they were centered on the maximum of the broad
band emission, and by aligning our broad band images with the [OIII]
image using two stars on the frames. Previously to any offsetting of
the slit, a short exposure on the nucleus with the same position angle
was performed, to allow the determination of the offset slit position.

\vspace{0.2truecm}

\begin{figure*}[ht!]
\centerline{\psfig{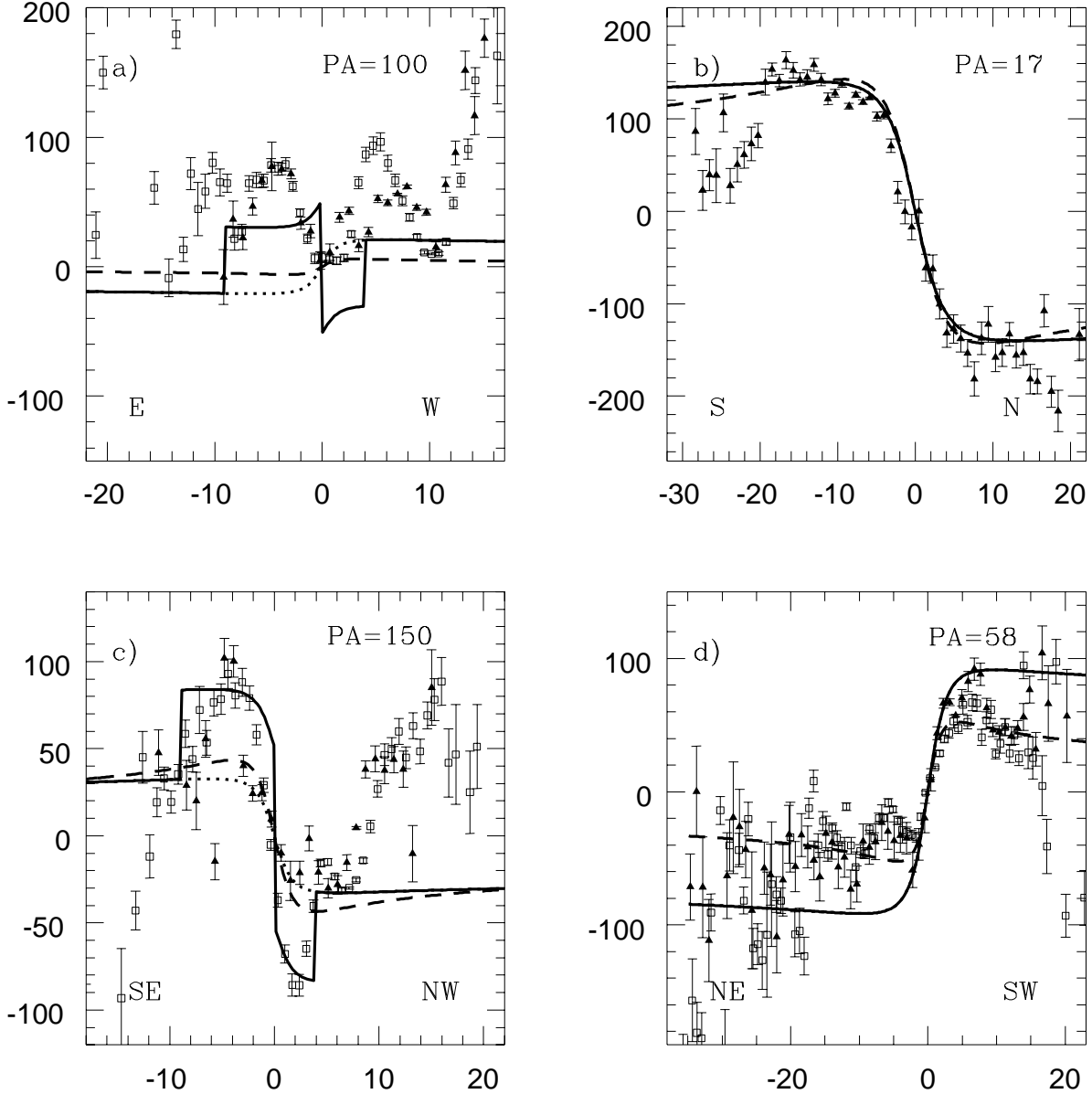}}
\caption[ ]{Velocities (in km s$^{-1}$) vs. distance to the nucleus
(in arcseconds) along the following slit position angles:
a) PA=100$^\circ$; b) PA=17$^\circ$; c) PA=150$^\circ$;
d) PA=58$^\circ$ (all these slits cross the nucleus). Squares
represent velocities in the [OIII] lines and triangles those in the
H$\alpha$ wavelength range. The dashed lines correspond to the first
model (rotation with $\phi=17^\circ$, $V_0$=350 km s$^{-1}$,
$r_0$=5 arcsec, $p$=1.3), the dotted lines to the second model (rotation with
$\phi$=30$^\circ$, $V_0$=250 km s$^{-1}$, $r_0$=5 arcsec, $p$=1.1). The full
lines correspond to the rotation model 2 plus a radial outflow with
$V$=150 km s$^{-1}$.}  \protect\label{velfield}
\end{figure*}

\begin{figure*}
\centerline{\psfig{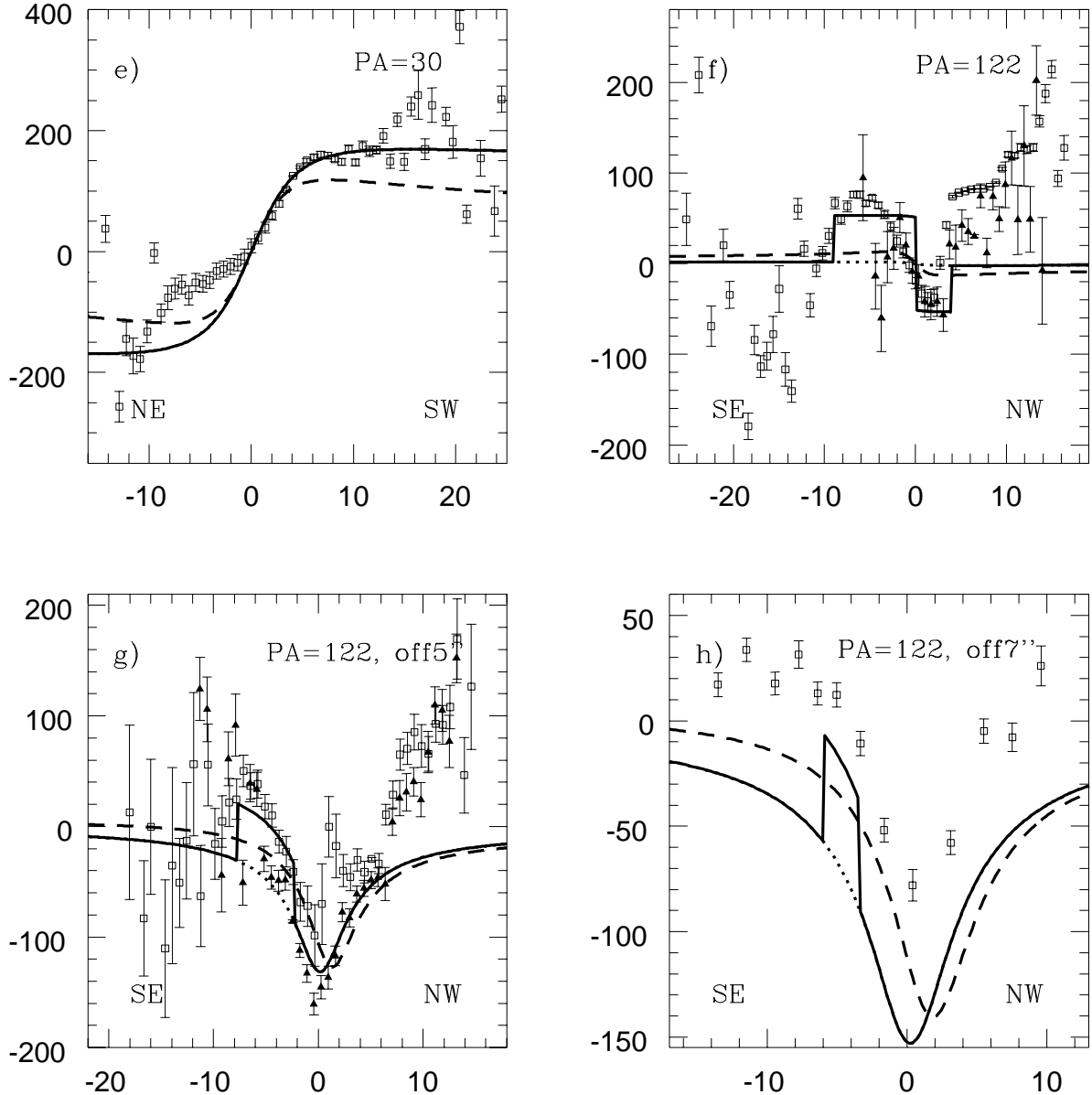}}
\caption[h]{Same as Fig. 2, along the following slit position angles:
e) PA=30$^\circ$; 
f) PA=122$^\circ$; g) PA=122$^\circ$, offset by 5 arcsec to the northeast; 
h) PA=122$^\circ$, 
offset by 7 arcsec to the northeast. }
\protect\label{velfield2}
\end{figure*}    

The data reduction was performed, following the standard procedures
for de-biasing, flat-fielding and calibrating the spectra, with the
IHAP and MIDAS softwares for the 1988 and 1991 runs respectively.

The accuracy of the wavelength calibration was checked by measuring
the position of the strong sky lines along each spectrum; the
dispersion of the sky line wavelengths are always within the error
bars of the calibration with the arc spectra. It is only for the
[OIII] spectra along PA=122/0 that we found a shift of 54 km s$^{-1}$,
which was applied to the final data for this PA. Typical wavelength
calibration errors give velocity uncertainties of $\pm$ 20 and $\pm$
60 km s$^{-1}$ for the 1988 and 1991 runs respectively.

Since the illumination of the slit in the considered spatial region is
uniform, a 2D-sky could be determined by averaging two strips on
either side off the galaxy and subtracted to the galaxy spectrum.

The ionized gas velocity field along each slit is obtained by applying
a cross-correlation method (Tonry \& Davis 1979) to the emission lines
by using a program developed by J. Perea within the FIGARO software;
this allows to calculate relative velocities with respect to a
reference cross-section chosen to have a high signal to noise ratio.
The redshift corresponding to this cross-section is measured by
fitting gaussian profiles to the different emission lines. We checked
that the cross sections corresponding to the nucleus on the various
slits had the same velocity within the error bars.  The systemic
velocity deduced is 2330 km s$^{-1}$, in agreement with the stellar
value measured by Nelson \& Whittle (1995).  Since the
cross-correlation method takes into account all the emission features
available in a given wavelength range, the velocities obtained are
weighted averages, and are therefore more representative of those of
the strongest lines, and/or strongest component in a line. Thus, as
velocities in the H$\beta$-[OIII] range are dominated by [OIII], the
high ionization component, we made an attempt to derive [OIII] and
H$\beta$ velocities separately. Unfortunately, H$\beta$ is only bright
enough close to the nucleus. We therefore decided to derive the
velocity field in [OIII] on one hand (hereafter the high excitation
domain), and in H$\alpha$-[NII] on the other (low excitation domain).
In any case, these are not peak velocities.  We plot the [OIII]
(squares) and H$\alpha$ (triangles) velocities in
Figs. \ref{velfield}-\ref{velfield2}-\ref{velfield3}.

The use of cross-correlation methods gives smaller uncertainties for
the relative velocity distributions than the typical wavelength
calibration errors.  This can be seen in
Figs. \ref{velfield}-\ref{velfield2}-\ref{velfield3}, where the error
bars correspond only to the velocity accuracy with respect to the
reference cross section. Therefore, realistic error bars are somewhat
larger.

Along PA=122$^\circ$ offset by 7 arcsec to the northeast (Fig. 3h), the
signal to noise was not sufficient and the cross--correlation method
could not be used. So we binned spatially the data over a few cross
sections and fitted gaussians to the line profiles to estimate
velocities.

Since the H$\alpha$ and H$\beta$ domains were not observed
simultaneously and the nights were not photometric, it is not possible
to draw excitation maps.

Notice that since the velocities derived from the cross correlation
are not peak velocities, the presence of various components in a
single line or different line shapes due to dust contamination
(producing asymmetric emission lines) can produce artificial
kinematical shifts (see below).

\section{The kinematical properties}

\begin{figure}[h!]
\centerline{\psfig{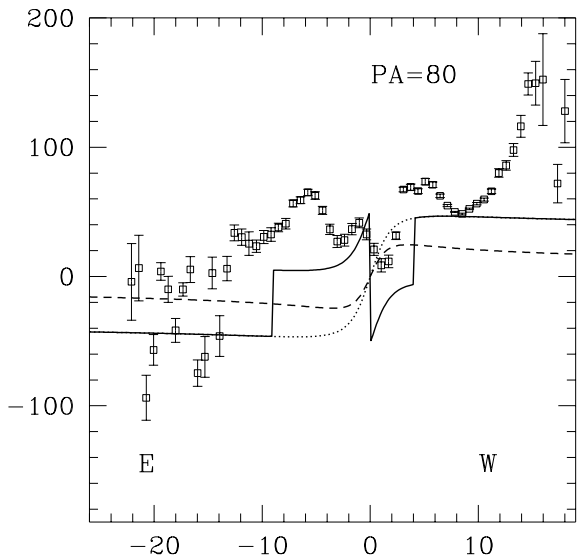}}
\caption[t]{Same as figure 2, along  PA=80$^\circ$. }
\protect\label{velfield3}
\end{figure}

Figures \ref{velfield}-\ref{velfield2}-\ref{velfield3} show the
velocity fields observed for a total of 14 long slit spectra in the
[OIII] and H$\alpha$ wavelength ranges along nine different PAs.  The
velocities are given relatively to the systemic velocity and plotted
as a function of distance to the nucleus for slits crossing the
nucleus.  For offset slits, the zero value in abscissa corresponds to
the minimum distance to the nucleus. The velocities plotted in
Figs. \ref{velfield}-\ref{velfield2}-\ref{velfield3} reveal a rather
complex velocity field which obviously cannot simply be due to
rotation of a disc with a major axis $\phi \simeq$15$^\circ$ (the
photometric major axis, RC3). Notice that our data are in very good
agreement with those of Heckman et al. (1981) along PA=120$^\circ$ and
of Keel (1996) along PA=16$^\circ$.  

Notice also that the high and low excitation gases do not always have
the same velocity structure.

\begin{figure}[h!]
\centerline{\psfig{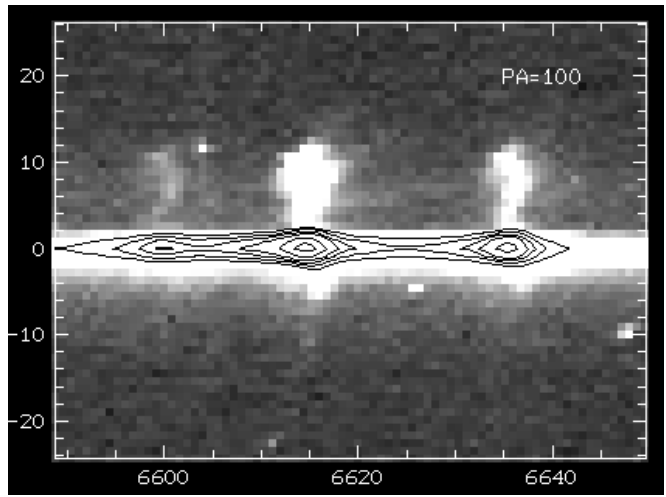}}
\caption[t]{Two-dimensional spectrum in the H$\alpha$-[NII] region 
along PA=100$^\circ$, showing complex line 
structure. East  is to the bottom and west to the top. The contours 
corresponding to the innermost regions have been superimposed.} 
\protect\label{pa280}
\end{figure}

\begin{figure}[h!]
\centerline{\psfig{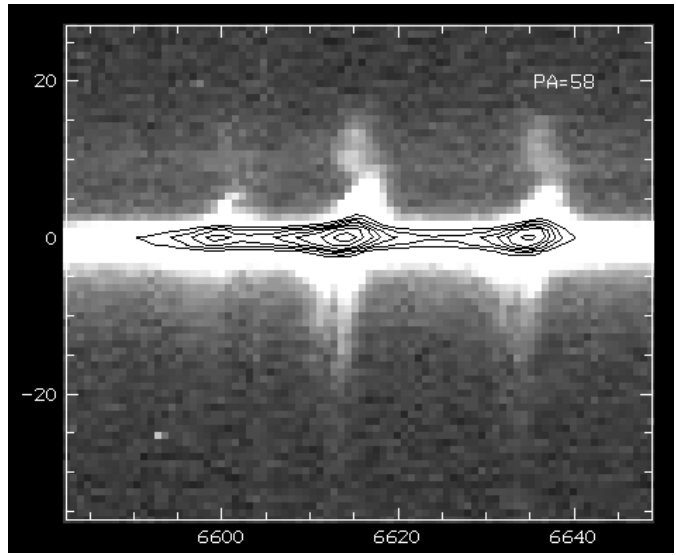}}
\caption[t]{Two-dimensional spectrum in the H$\alpha$-[NII] region
along PA=58$^\circ$ showing clear line splitting. Northeast is to the bottom 
and south-west to the top. The contours corresponding to the innermost 
regions have been superimposed. }
\protect\label{pa238}
\end{figure}

\subsection{Line profiles} 

The complexity of the velocity field is illustrated in two 2D plots of
the high resolution spectra (Figs. \ref{pa280} and  \ref{pa238}).  A
blue asymmetry of the lines is detected in the central 3 arcsec (see
Figs. \ref{pa280} and  \ref{pa238}), as already noted by Colina et
al. (1987). On the west side, the velocity dispersion can be seen to
be very large e.g. along PA=100$^\circ$ (see Fig. \ref{pa280}) giving
a ``mushroom'' shaped spectrum with a typical FWHM of 300 km s$^{-1}$.

Although in the [OIII] image almost no emission is seen to the north
and east of the nucleus out to 5 arcsec, this is probably due to an
oversubtraction of the continuum. However, weak emission is detected
in our spectra (see for example Figs. 5 and 6).

\begin{figure}[h!]
\centerline{\psfig{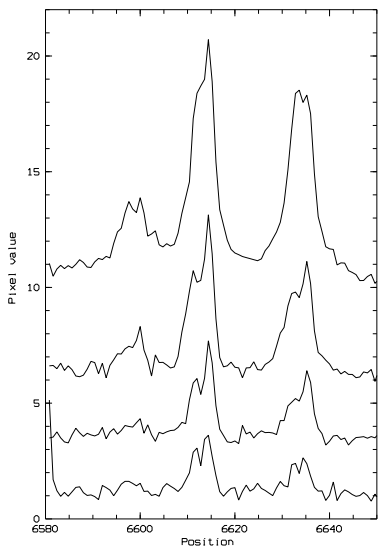}}
\caption[ ]{Spectra in the H$\alpha$-[NII] region 
of several cross-sections along PA=58$^\circ$ showing 
complex line profile structure. The top spectrum is 2.3 arcsec northeast of the 
nucleus, and the following spectra from top to bottom are drawn every 1.8 arcsec
towards the northeast.}
\protect\label{onedspec}
\end{figure}

The high spectral resolution H$\alpha$+[NII] data reveal the existence
of double peaked profiles in the east quadrant, along PA=17$^\circ$ in
the south, PA=58$^\circ$ in the northeast and PA=150$^\circ$ in the
southeast.  These regions extend over $\sim$10 arcsec.  The most striking
example is given Fig. \ref{pa238} along PA=58$^\circ$.  In order to
estimate the velocities of these two components, we performed a
synthesis analysis using multiple gaussian fitting along this PA; the
respective typical FWHMs of the two components are 90 and
250 km s$^{-1}$ on the northeast side, at a distance of about 5 arcsec from
the nucleus. Both components (estimated at about $\sim$5 arcsec from the
nucleus) are blueshifted by $-135$ km s$^{-1}$ and $-10$ km s$^{-1}$
with respect to the systemic velocity. Examples of one-dimensional
spectra are shown in Fig. \ref{onedspec}. Along PA=150$^\circ$ and at
5 arcsec from the nucleus, the two components are redshifted by 135 and
30 km s$^{-1}$. The fact that line splitting is observed only for the
low excitation gas does not mean that the gas producing such features
is less ionized, but may merely be due to the lower spectral
resolution of our [OIII] data.

Such line splitting probably represents radial motions (outflow or
inflow) of the gas in the NLR, a rather common feature of extended
high excitation gas in Seyferts (see Christopoulou et al., 1997;
Lindblad et al. 1996; Morris et al., 1985; Storchi-Bergmann et al., 1992; 
Wilson et al., 1985).  In fact, the morphology of the emission region is
reminiscent of a cone (Fig. \ref{slits}, and Fig. 4 in Wehrle \&
Morris, 1988) although the possibility that the arc is a ring or
spiral structure around the nucleus cannot be discarded.

\subsection{Velocity field}

In order to disentangle overall rotation from peculiar motion, we
first apply a simple galactic rotation model. The gas is considered in
circular motion and lies in a disc with an inclination $i=70^\circ$
and a position of the major axis $\phi=15^\circ$ (RC3).

The velocity law, which assumes that the gas is in a spherical
gravitational potential and follows circular orbits in a plane (de
Zeeuw \& Lynden-Bell 1988) is given by the following equation:

$$V(r)-V_{syst} = \frac{V_0\; r}{(r^2+r_0^2)^{p/2}}$$

\noindent
where $r$ is the distance to the center in the plane of the galaxy,
$V_0$ the maximum amplitude and $r_0$ the distance at which this
velocity is reached; $p$ is a parameter controlling the slope of the
rotation curve in its ``flat'' outer regions.

We assumed that the dynamical center coincides with the peak of the
continuum light and constrained the above parameters using the velocity 
field observed along the photometric major axis, i.e. $\phi=17^\circ$.
The resulting rotation curve computation can be seen as a dashed line
in Figs. \ref{velfield}-\ref{velfield2}-\ref{velfield3}.  We tried
different sets of parameters and obtain a fairly good match for
$V_0$=350 km s$^{-1}$, $r_0$=5 arcsec and $p$=1.3. Indeed, as mentioned
above, $p$ cannot be much different, since it is constrained by the
shape of the rotation curve beyond 5 arcsec; $r_0$ could vary between 5 arcsec
and 8 arcsec, and $V_0$ by no more than $\pm 10$ km s$^{-1}$.

However it is apparent on Figs. 3g and 3h, that $\phi=17^\circ$ is
not the best estimate for the kinematic major axis. Indeed, the
parameters lead to velocity fields which are systematically shifted
with respect to the measurements along PA=122$^\circ$
(Figs. 3g-h). Along PA=30$^\circ$ (Fig. 3e), the velocity field which
becomes flat at distances over 5 arcsec is badly represented. The steep
central gradient and the amplitude of flat rotation regions further
out indeed suggest that the kinematical major axis of the ionized gas
is larger than $\phi$=17$^\circ$.

This value is strongly constrained by the PA=122$^\circ$ offset data
(Figs. 3g-h), in order to shift the minimum model velocity to the
centre, and can only vary between 25 and 30$^\circ$. The set of
parameters is somewhat different in this case: $V_0$=250 km s$^{-1}$
and $p$=1.1, the maximum velocity being reached at the same radius,
$r_0$=5 arcsec.  The resulting rotation velocity field is plotted in
Figs. \ref{velfield}-\ref{velfield2}-\ref{velfield3} as a dotted line.

Better matches to the data in most of the slit positions are obviously
obtained. The fact that the photometric major axis derived from the
kinematics does not agree with that derived from broad band imaging
suggests that the gas disk is in a different plane than the stellar
disk (for instance because of a warping due to the interaction with
NGC 2993) and/or that a radial component is present in addition to the
galactic rotation. It is worth noticing that PA=30$^\circ$ is
perpendicular to the PA=130$^\circ$ of the elongated radio structure
as measured by Hummel et al. (1983) from VLA observations. This model
accounts for most of the low excitation gas structure along non offset
slits in the central few arcseconds (excepting PA=150$^\circ$);
however, further from the nucleus, the low and high excitation gases
do not always follow the same kinematics.

As discussed in Sect. 3.1, outflow of the gas within a conical
envelope or on the surface of a hollow cone is a possible picture to
account for the double peaks and blueshift.  The morphology of the
emitting gas (see the [OIII] image in Fig. 1 and Fig. 4 in Wehrle \&
Morris 1988) is suggestive of a cone structure along an axis at
PA=120$^\circ$ and with a projected full opening angle of about
120$^\circ$.  Notice that the cone axis is almost aligned with that of
the elongated radio structure observed by Hummel et al. (1983) with
the VLA, and its linear extent is comparable to that of the radio
emission, namely 20 arcsec on the east side and $\leq$10 arcsec on the west
side.

Although the line splitting is ignored when measuring the H$\alpha$
lines, two components are clearly present in the low excitation gas
along PA=58$^\circ$ and 150$^\circ$ (see Fig. 7 and Sect. 3.1). One
can notice in Fig. 2c that +30 km s$^{-1}$, the velocity of one
component along PA150$^\circ$, is close to the velocity expected from
a pure disk rotation (dashed curve). In Fig 2d, the value of
-10 km s$^{-1}$ is not far from the disk rotation velocity.  It is
thus clear that the plotted weighted mean velocities are dominated by
outflow in the east.  We have therefore used a very simple model in
which the outflowing velocities are considered as radial motions in a
plane very close to that of the gas disk. A more sophisticated
modeling of a true cone, such as that of Wilson et al.  (1985), Hjelm
\& Lindblad (1996a, 1996b) or Christopoulou et al.  (1997) is not
possible due to the lack of detailed information on the line splitting
and of photometrical information.

We have considered that the west side of the galaxy is the near one;
this would explain why the emission cone is seen mainly to the east.
We add to the general rotation pattern described above a constant
outflow, modeled along a triangular region of axis PA=120$^\circ$,
with an opening angle of 120$^\circ$ and a constant velocity of
150 km s$^{-1}$ along the outflow region (measured 5 arcsec from the
nucleus). This is a simple addition of outflowing radial velocity in a
region determined by the [OIII] image morphology and taking place in a
plane very close to the gas disk.  The projected size of the zone on
which this region extends its influence is taken to be 9 arcsec and 4 arcsec in
the east and west regions respectively; these values are strongly
constrained by the high excitation gas data along PAs between
30$^\circ$ and 150$^\circ$ (centered and offcentered).  The results
are shown as full lines in
Figs. \ref{velfield}-\ref{velfield2}-\ref{velfield3}.  Although this
model is very simple, it accounts notably better than the previous one
for the high excitation gas, as can be seen in Figs. 2c, 3e, 3f and
3g. Nevertheless, close inspection of
Figs. \ref{velfield}-\ref{velfield2}-\ref{velfield3} raises some
comments.

The low excitation gas represented by the lines in the H$\alpha$
domain follows more closely the pattern of normal rotation with a
kinematical major axis $\phi$=30$^\circ$. Notice that along
PA=17$^\circ$ the H$\alpha$ data follows a rotation pattern with
$\phi$=30$^\circ$; the discrepant points in the SW might correspond to
the annulus that is conspicuous at such distances on broad band
images; it could correspond to part of a spiral arm or ring, in which
non-circular motions are expected.

For the whole set of PAs, the data for both the high and low
excitation gas at distances larger than 10 arcsec to the northwest show the
largest differences with respect to the model.  This could suggest
that we are seeing kinematically distinct regions, i.e. line of sight
gas which is not physically associated with the main emission
structure. Indeed, the gas kinematics could be perturbed by the
interaction of NGC 2992 with its close companion NGC 2993; notice that
the existence of a warp in the gaseous disk is suggested by the
difference between the photometrical and kinematical major axes.

Another feature that can perturb the measured velocity field is dust.
Indeed, if the line profiles are affected by dust, the
cross-correlation method used to infer velocity points will result in
velocities which are not fully characteristic of the sampled region.
This could explain the humps observed to the SE along PA=150$^\circ$ and
100$^\circ$, and also account for the differences observed between the
[OIII] and H$\alpha$ ranges in the NW along PA=100$^\circ$.

Along PA=122$^\circ$ offset by 5 arcsec (Fig. 3g), the hump observed in the
velocity field just southeast of the nucleus corresponds to the
annulus of ionized gas already mentioned above, while the large values
northwest of the nucleus correspond to the ``finger'' observed by
Wehrle \& Morris (1988). It is probable therefore that these two
regions do not follow the general rotation pattern. Notice that, here
again, H$\alpha$ roughly follows the disc rotation while the [OIII]
velocities are better accounted for by a model including outflow.

\section{Conclusions}

We have obtained the gas kinematics of NGC 2992 by means of long slit
data along nine position angles. The results are in general agreement
with previous determinations (Heckman et al. 1981, who presented data
along PA=120$^\circ$ with higher spectral resolution but lower spatial
resolution; Colina et al. 1987, who presented data of lower spectral
and spatial resolution along 3 PAs; Keel 1996, for data with a
resolution similar to ours along PA$\simeq 17^\circ$).
  
We have modeled the kinematics of NGC 2992 by circular rotation in a
gaseous disk to which is added constant radial outflow in the disk
plane, as suggested by line splitting. Disk rotation can be accounted
for with the following parameters: major axis along $\phi$=30$^\circ$,
inclination $i$=70$^\circ$, velocity amplitude $V_0$=250 km s$^{-1}$,
and parameters $p$=1.1 and $r_0$=5 arcsec (see Sect. 3). This value of
$\phi$, which allows a fairly good representation of the observed
velocity field in the central regions, results to be different from
that of the large scale disk as derived by continuum images. This
discrepancy could imply that the disk of NGC 2992 is warped, probably
due to interaction with NGC 2993. Outflow was modeled along a
triangular region of axis PA=120$^\circ$, with an opening angle of
120$^\circ$ and a constant velocity of 150 km s$^{-1}$ along the
outflow region (measured 5 arcsec from the nucleus). Opening the angle of
the outflow to 160$^\circ$ in the east could give a better fit only
for PA=30$^\circ$ in the northeast, in the region where H$\alpha$
emission is detected by Wehrle \& Morris (1988).

It can be noted that low excitation gas follows better the pure
rotation model in some regions, whereas high excitation gas is better
represented when outflow is also included in the model.  In our
scenario, the outflow takes place radially close to the plane of the
gaseous disk, with a spatial extension which is much larger on the
east than on the west side.

This simple model accounts rather well for the kinematics of NGC 2992,
if one excepts the regions located more than 6 arcsec northwest of the
nucleus. The kinematics of the gas in that region confirm results by
Heckman et al. (1981) and Colina et al. (1987), who reported large
velocities and velocity dispersions. As noticed before, these regions
with the largest discrepancies may be understood in terms of more
complex dynamics and structure, and/or by taking into account the
possible line asymmetries due to the dust distribution. Since our
spectral resolution is not sufficient to observe line splitting in
high excitation lines, and we don't have line ratios, it could be
difficult to constrain a more sophisticated model.

\begin{acknowledgements}

We acknowledge the referee, P.O. Lindblad, whose suggestions helped us
to improve the paper considerably.  We are very grateful to F. Warin
for her help in the data reduction, to J. Perea for making his
cross-correlation program available to us and to D. Pelat for giving
us his profile fitting programme.  I. M\'arquez acknowledges financial
support from the Spanish Ministerio de Educaci\'on y Ciencia.

\end{acknowledgements}

\end{document}